\documentclass[a4paper,conference]{IEEEtran}
\ifCLASSOPTIONcompsoc
\usepackage[nocompress]{cite}
\else
\usepackage{cite}
\fi
\usepackage{amsmath}

\usepackage{graphicx}

\ifCLASSINFOpdf
\else
\fi
\usepackage{tikz}
\usetikzlibrary{positioning}
\usepackage[hyphens]{url}

\begin{document}
%
\title{Image Super-Resolution Quality Assessment: Structural Fidelity Versus Statistical Naturalness\vspace{-0.8cm}}
\author{\IEEEauthorblockN{Wei Zhou$^{1,2}$, Zhou Wang$^{1}$, Zhibo Chen$^{2}$ \\
$^{1}$Dept. of Electrical \& Computer Engineering, University of Waterloo, Waterloo, ON N2L3G1, Canada \\
$^{2}$CAS Key Laboratory of Technology in Geo-spatial Information Processing and Application System \\
University of Science and Technology of China, Hefei 230027, China \\
Email: \{wei.zhou, zhou.wang\}@uwaterloo.ca; chenzhibo@ustc.edu.cn}}


%



\maketitle

\begin{abstract}
Single image super-resolution (SISR) algorithms reconstruct high-resolution (HR) images with their low-resolution (LR) counterparts. It is desirable to develop image quality assessment (IQA) methods that can not only evaluate and compare SISR algorithms, but also guide their future development. In this paper, we assess the quality of SISR generated images in a two-dimensional (2D) space of structural fidelity versus statistical naturalness. This allows us to observe the behaviors of different SISR algorithms as a tradeoff in the 2D space. Specifically, SISR methods are traditionally designed to achieve high structural fidelity but often sacrifice statistical naturalness, while recent generative adversarial network (GAN) based algorithms tend to create more natural-looking results but lose significantly on structural fidelity. Furthermore, such a 2D evaluation can be easily fused to a scalar quality prediction. Interestingly, we find that a simple linear combination of a straightforward local structural fidelity and a global statistical naturalness measures produce surprisingly accurate predictions of SISR image quality when tested using public subject-rated SISR image datasets. Code of the proposed SFSN model is publicly available at
\url{https://github.com/weizhou-geek/SFSN}.
\end{abstract}

\begin{tikzpicture}[overlay, remember picture]
\path (current page.north) node (anchor) {};
\node [below=of anchor] {%
2021 Thirteenth International Conference on Quality of Multimedia Experience (QoMEX)};
\end{tikzpicture}
%


%
\IEEEpeerreviewmaketitle

\vspace{-0.8cm}
\section{Introduction}
Single image super-resolution (SISR) aims to recover a high-resolution (HR) image given a single low-resolution (LR) image. SISR plays a significant role in a wide range of applications, from satellite imaging, web browsing, to video surveillance \cite{park2003super}. During the past decades, numerous SISR algorithms have been proposed, including interpolation-based \cite{keys1981cubic,wang2007new}, dictionary-based \cite{yang2010image,wang2012semi,yang2016consistent}, and deep learning-based methods \cite{kim2016accurate,johnson2016perceptual,yamanaka2017fast,tai2017image,lai2017deep,ledig2017photo}. The visual appearance and quality of SISR generated images vary dramatically when different SISR approaches are employed. Nevertheless, there is still no consensus on how the quality of SISR created images should be assessed. This is critically important because image quality assessment (IQA) methods not only help evaluate and compare SISR algorithms, but also guide the development of future SISR methodologies.

In general, the most reliable quality assessment method is human subjective evaluation \cite{yeganeh2015objective,ma2017learning,zhou2019visual}. But subjective tests are usually expensive, time-consuming and hard to be integrated into SISR optimization frameworks. Therefore, it is highly desirable to design effective objective IQA models for SISR generated images. Depending on the availability of the original pristine image, full-reference (FR) IQA \cite{wang2004image,wang2003multiscale,zhang2011fsim,sampat2009complex,liu2011image,xue2013gradient,sun2018spsim} and no-reference (NR) IQA approaches \cite{mittal2012no,mittal2012making,saad2012blind,moorthy2011blind,wu2015highly} may be applied. Additionally, since many SISR algorithms produce blurry reconstructed images, image sharpness assessment (ISA) or blur measures \cite{vu2011bf,hassen2013image,hosseini2019encoding} may also be employed.

\begin{figure}[t]
  \centering
  \includegraphics[width=0.35\textwidth]{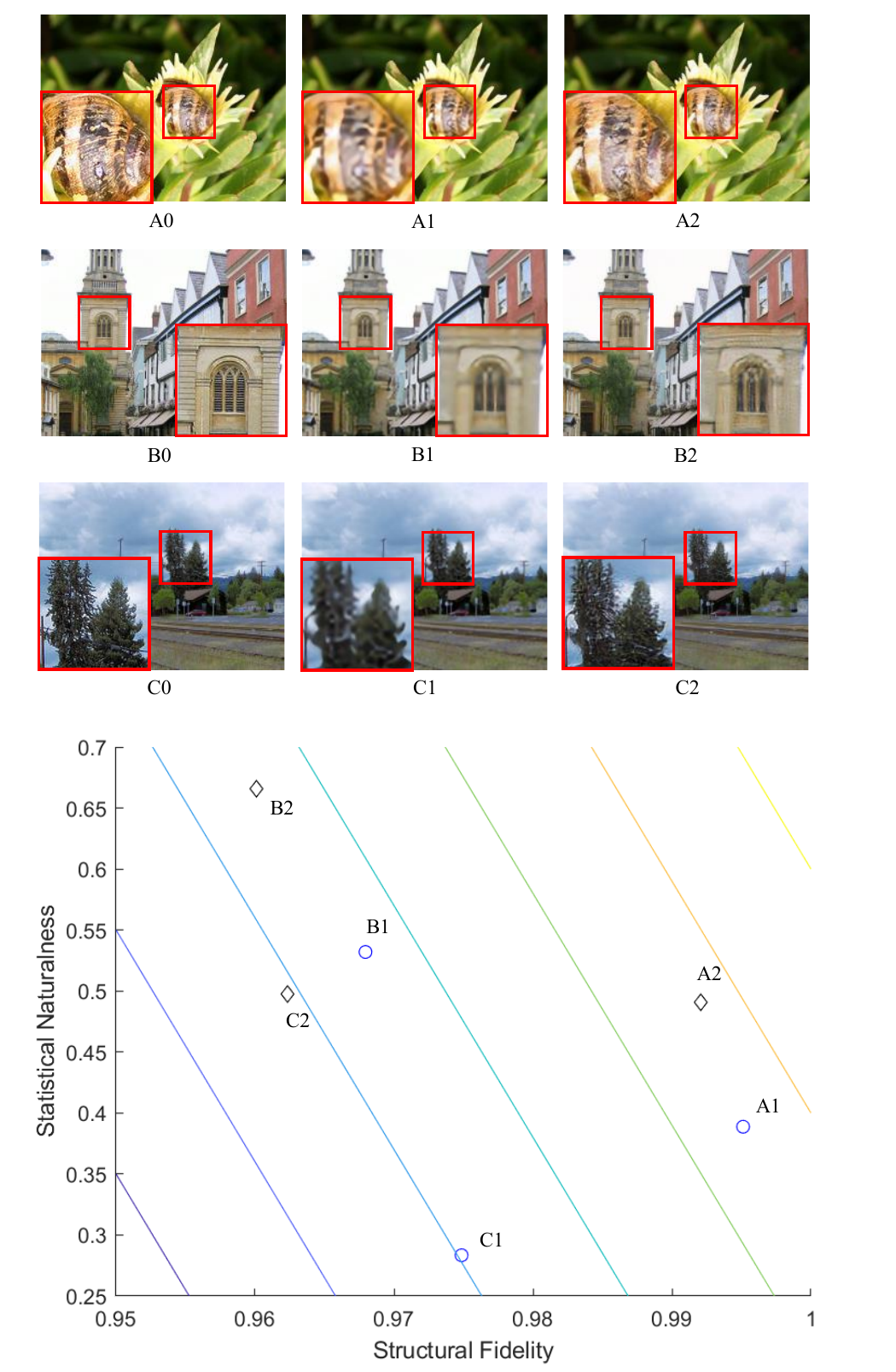}
  \caption{SISR generated images and 2D quality assessment of statistical naturalness versus structural fidelity ({\it SF} vs {\it SN}). A0, B0, C0: original HR images; A1, B1, C1: reconstructed images by VDSR~\cite{kim2016accurate} at scaling factor 4; A2, B2, C2: reconstructed images by SRGAN~\cite{ledig2017photo} at scaling factor 4.}
  \label{fig1}
\end{figure}

\begin{figure*}[t]
  \centering
  \includegraphics[width=0.70\textwidth]{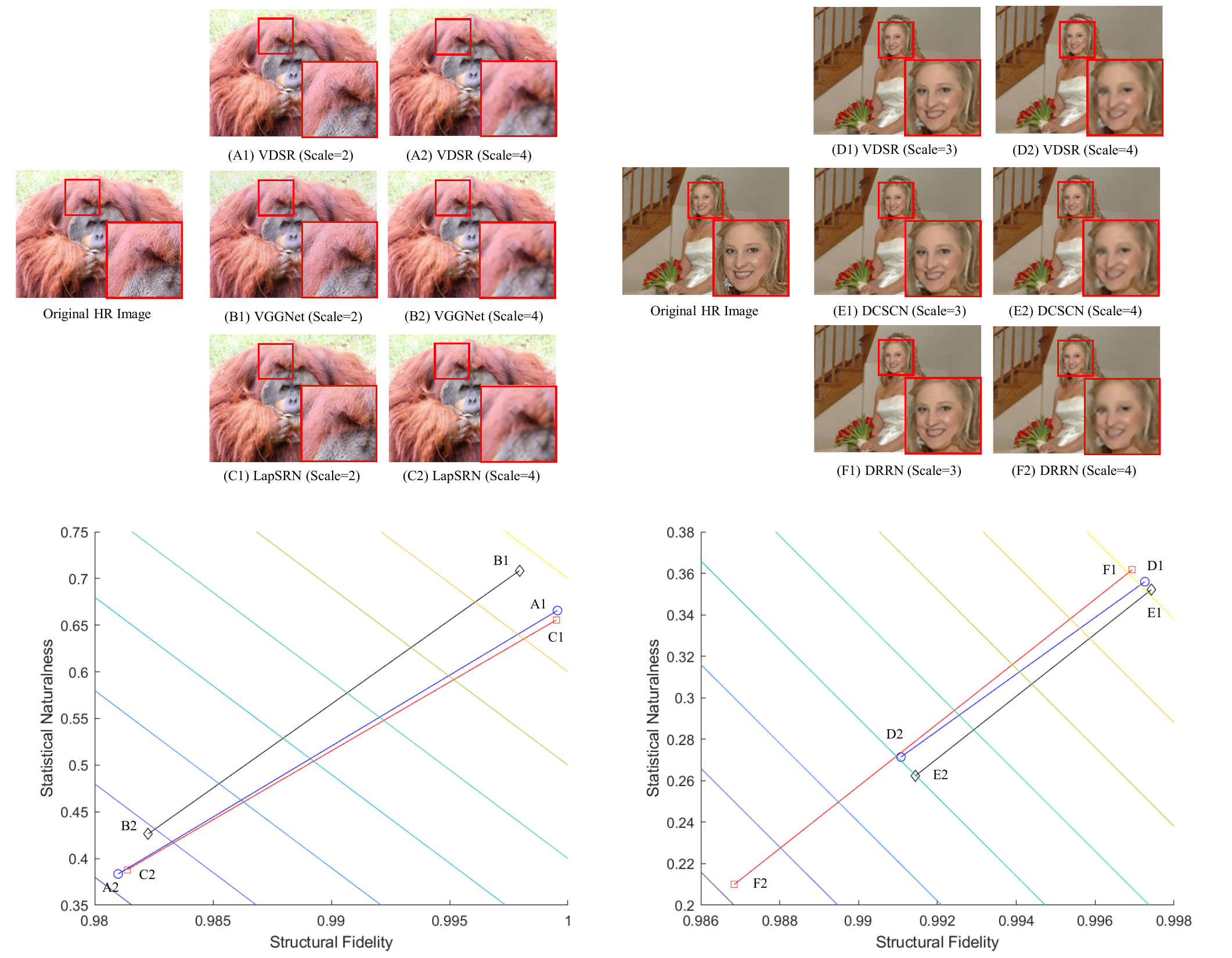}
  \caption{Sample SISR generated images at different scaling factors (top) and their corresponding points in ({\it SF} vs {\it SN}) plots (bottom), where the level sets of the proposed quality prediction $Q({\bf y})$ are also shown.}
  \label{fig2}
\end{figure*}

Despite the success in other IQA applications, existing FR-IQA, NR-IQA and ISA methods often fall short when evaluating the quality of SISR generated images. The gap is not only on the accuracy in predicting subjective scores, but also on effectively interpreting the nature of key quality degradation trends in SISR images. An example is given in Fig.~\ref{fig1}, where traditional SISR methods such as VDSR~\cite{kim2016accurate} are highly effective at achieving high signal fidelity in terms of signal-to-noise ratio or structural similarity \cite{wang2004image} when compared to the original images, but the resulting images often look artificial. On the other hand, recently proposed generative adversarial network (GAN) based approaches such as SRGAN~\cite{ledig2017photo} are impressive at producing natural-looking reconstructed images, but their signal fidelity measures are significantly lower. These observations motivate us to look at the problem in a two-dimensional (2D) space of structural fidelity versus statistical naturalness, as demonstrated at the bottom part of Fig.~\ref{fig1}.




\section{2D Quality Assessment of SISR Images}




Multi-scale image decomposition such as Laplace and wavelet transforms have been shown to be  effective at characterizing not only local perceptual degradation, but also the statistical naturalness of images. Therefore, we apply a multi-scale image transform and construct local structural fidelity and global statistical naturalness measures both in the transform domain. Given the original HR image ${\bf x}$, and the SISR generated test image ${\bf y}$,  inspired by the success of MS-SSIM~\cite{wang2003multiscale}, we define subband patch level structural fidelity measure as:
\begin{equation}\label{5}
\centering
SF_{local}^k(x,y)=\frac{{{\sigma }_{xy}}+C}{{{\sigma }_{x}}{{\sigma }_{y}}+C},
\end{equation}
where $x$ and $y$ denote the patches extracted from the $k$-th subband from ${\bf x}$ and ${\bf y}$, respectively, ${{\sigma }_{x}}$ and ${{\sigma }_{y}}$ are their standard deviations, ${{\sigma }_{xy}}$ represents the covariance between $x$ and $y$, and $C$ is a positive stabilizing constant. The scale level structural fidelity measure is then computed by spatial pooling:
\begin{equation}\label{6}
\centering
SF^k({\bf x},{\bf y})=\frac{1}{M}\sum\limits_{m=1}^{M}{SF_{local}^k(x,y)},
\end{equation}
where $M$ denotes the number of local patches in the subband. Finally, we fuse across scales to obtain the overall structural fidelity between ${\bf x}$ and ${\bf y}$:
\begin{equation}\label{7}
\centering
SF({\bf x},{\bf y})=\prod\limits_{k=1}^{K}[{SF^k}({\bf x},{\bf y})]^{{\alpha }_k},
\end{equation}
where $K$ is the total number of scales/subbands, and ${{\alpha }_{k}}$ is the weight assigned to the $k$-th scale as in \cite{wang2003multiscale}. Furthermore, natural texture-rich content tends to have higher entropy in the transform domain~\cite{chen2017blind}. Thus, we use the global entropy of transform coefficients as a statistical naturalness measure:
\begin{equation}\label{8}
\centering
SN({\bf y})=-\sum{P({c_y})\log (P({c_y}))},
\end{equation}
where $P({c_y})$ denotes the probability of subband coefficients of the test image ${\bf y}$ and may be approximated with histograms.

Although the proposed pair of ({\it SF}, {\it SN}) measure is rather simple, it offers a meaningful 2D illustration of the behaviors of SISR algorithms. Fig.~\ref{fig1} shows three original HR images with their corresponding SISR images generated by VDSR~\cite{kim2016accurate} and SRGAN~\cite{ledig2017photo}. The ({\it SF} vs. {\it SN}) plot clearly indicates the relative advantage of VDSR over SRGAN on the {\it SF} measure, and conversely the advantage of SRGAN over VDSR on the {\it SN} measure. The pattern is consistent over all three content, as indicated by the pairs of points (A1, A2), (B1, B2) and (C1, C2). Fig.~\ref{fig2} shows images generated by different SISR algorithms applied to LR images of different sizes and enhanced by different scaling factors. The ({\it SF} vs. {\it SN}) plots offer a platform to examine the behaviors of different SISR algorithms across scaling factors. It can be observed that the the general trend of any SISR algorithm is that both {\it SF} and {\it SN} measures drop with increasing scaling factors. However, the speed of change may vary depending on the algorithm and possibly the image content. For example, the DRRN~\cite{tai2017image} method appears to be much more sensitive to scaling factor change than VDSR~\cite{kim2016accurate} and DCSCN~\cite{yamanaka2017fast} for the right image.


\section{Fusing 2D Assessment for 1D Prediction}
In practice, it is often desirable to obtain a single quality score indicating the overall quality of SISR generated images. This can be achieved by collapsing the proposed 2D measure into a scalar quality prediction, e.g., by a linear combination:
\begin{equation}\label{9}
\centering
Q({\bf y})=w_F SF({\bf x},{\bf y})+w_N SN({\bf y}),
\end{equation}
where the weighting factors $w_F$ and $w_N$ adjust the relative importance of the two measures, and are set empirically at 0.9 and 0.1, respectively, in the current implementation. We name $Q$ the SFSN measure, which creates straight lines as level sets in the 2D space, as shown in the bottom plots of Figs.~\ref{fig1} and \ref{fig2}. This scalar quality prediction can then be validated by comparing against subject-ratings of SISR generated images.

\renewcommand\arraystretch{0.8}
\begin{table}[h]
	\centering
	\scriptsize
	\caption{SRCC Performance comparison of objective models on WIND \cite{yeganeh2015objective}, CVIU \cite{ma2017learning} and QADS \cite{zhou2019visual} databases.}
	\begin{tabular}{c|c|c|c|c}
		\hline
		\textbf{Methods} & \multicolumn{1}{c|}{\textbf{WIND}} & \multicolumn{1}{c|}{\textbf{CVIU}} & \multicolumn{1}{c|}{\textbf{QADS}} & \multicolumn{1}{c}{\textbf{Average}} \\ \hline
		\textbf{PSNR} &0.6320 &0.5663 &0.3544 &0.5176 \\
		\textbf{SSIM \cite{wang2004image}} &0.6125 &0.6285 &0.5290 &0.5900 \\
		\textbf{MS-SSIM \cite{wang2003multiscale}} &0.8246 &0.8048 &0.7172 &0.7822 \\
		\textbf{FSIM \cite{zhang2011fsim}} &0.8503 &0.7481 &0.6885 &0.7623 \\
		\textbf{CW-SSIM \cite{sampat2009complex}} &0.8626 &0.7591 &0.3259 &0.6492 \\
        \textbf{GSIM \cite{liu2011image}} &0.7649 &0.6505 &0.5538 &0.6564 \\
        \textbf{GMSD \cite{xue2013gradient}} &0.7966 &0.8469 &0.7650 &0.8028 \\
		\textbf{SPSIM \cite{sun2018spsim}} &0.8141 &0.6698 &0.5751 &0.6863 \\ \hline
		\textbf{BRISQUE \cite{mittal2012no}} &0.7676 &0.5863 &0.5463 &0.6334 \\
        \textbf{NIQE \cite{mittal2012making}} &0.6263 &0.6525 &0.3977 &0.5588 \\
        \textbf{BLIINDS-II \cite{saad2012blind}} &0.5281 &0.3705 &0.3838 &0.4275 \\
        \textbf{DIIVINE \cite{moorthy2011blind}} &0.5465 &0.5479 &0.4817 &0.5254 \\
        \textbf{LPSI \cite{wu2015highly}} &0.6669 &0.4883 &0.4079 &0.5210 \\ \hline
		\textbf{S3 \cite{vu2011bf}} &0.4455 &0.5050 &0.4636 &0.4714 \\
		\textbf{LPC-SI \cite{hassen2013image}} &0.5375 &0.5450 &0.4902 &0.5242 \\
		\textbf{HVS-MaxPol-1 \cite{hosseini2019encoding}} &0.6166 &0.6421 &0.6170 &0.6252 \\
		\textbf{HVS-MaxPol-2 \cite{hosseini2019encoding}} &0.6309 &0.6313 &0.5736 &0.6119 \\ \hline
        \textbf{Proposed ({\it SF} only)} &0.8642 &0.8546 &0.7867 &0.8352 \\
        \textbf{Proposed ({\it SN} only)} &0.5873 &0.6415 &0.6115 &0.6134 \\
		\textbf{Proposed SFSN} &\textbf{0.8867} &\textbf{0.8714} &\textbf{0.8407} &\textbf{0.8663} \\ \hline
	\end{tabular}
\label{table1}
\end{table}

We validate the proposed fused SFSN quality prediction method on three public SISR IQA databases, including WIND \cite{yeganeh2015objective}, CVIU \cite{ma2017learning}, and QADS \cite{zhou2019visual}. The WIND database considers 8 interpolation algorithms with scaling factors of 2, 4, and 8. It contains 312 SISR images corresponding to 13 reference images. The CVIU database consists of 30 reference HR images and 1,620 SISR generated images created by 9 algorithms with 6 pairs of (scaling factor,  kernel width) combinations, where a larger scaling factor corresponds to a larger blur kernel width. The QADS database contains 20 original HR images and 980 images generated by 21 SISR algorithms, including 4 interpolation-based, 11 dictionary-based, and 6 deep learning (DL) based models applied for upsampling factors of 2, 3, and 4. In all three databases, each SISR generated image is subject-rated and annotated by a mean opinion score (MOS). We compare the proposed method with 8 FR-IQA, 5 NR-IQA, and 4 ISA models. The Spearman Rank-order Correlation Coefficient (SRCC) comparison results are reported in Table~\ref{table1}, where the best performances are highlighted in bold. Other common evaluation criteria~\cite{zhou2020blind} produce similar results but are not included due to space limit. Despite its simple and straightforward construction, SFSN achieves surprisingly competitive performance against state-of-the-art IQA and ISA models.

\renewcommand\arraystretch{0.8}
\begin{table}[h]
	\centering
	\scriptsize
	\caption{SRCC Performance comparison of objective models on different SISR categories on QADS~\cite{zhou2019visual} database.}
	\begin{tabular}{c|cccc}
		\hline
		\textbf{Methods} & \textbf{Interpolation} & \textbf{Dictionary} & \textbf{DL} & \textbf{Overall} \\ \hline
		\textbf{PSNR} &0.2972 &0.3808 &0.2656 &0.3544 \\
		\textbf{SSIM \cite{wang2004image}} &0.4015 &0.5481 &0.5121 &0.5290 \\
		\textbf{MS-SSIM \cite{wang2003multiscale}} &0.6340 &0.7425 &0.7104 &0.7172 \\
		\textbf{FSIM \cite{zhang2011fsim}} &0.5471 &0.6846 &0.6637 &0.6885 \\
		\textbf{CW-SSIM \cite{sampat2009complex}} &0.5254 &0.4362 &0.0986 &0.3259 \\
        \textbf{GSIM \cite{liu2011image}} &0.3946 &0.5332 &0.5661 &0.5538 \\
        \textbf{GMSD \cite{xue2013gradient}} &0.7054 &0.7709 &0.7363 &0.7650 \\
		\textbf{SPSIM \cite{sun2018spsim}} &0.4545 &0.5518 &0.5871 &0.5751 \\ \hline
		\textbf{BRISQUE \cite{mittal2012no}} &0.5096 &0.4951 &0.4357 &0.5463 \\
        \textbf{NIQE \cite{mittal2012making}} &0.4639 &0.4547 &0.4190 &0.3977 \\
        \textbf{BLIINDS-II \cite{saad2012blind}} &0.1814 &0.3628 &0.6547 &0.3838 \\
        \textbf{DIIVINE \cite{moorthy2011blind}} &0.4267 &0.4175 &0.5654 &0.4817 \\
        \textbf{LPSI \cite{wu2015highly}} &0.2726 &0.3309 &0.6034 &0.4079 \\ \hline
		\textbf{S3 \cite{vu2011bf}} &0.4016 &0.3171	&0.5458	&0.4636 \\
		\textbf{LPC-SI \cite{hassen2013image}} &0.3301 &0.3798 &0.2558 &0.4902 \\
		\textbf{HVS-MaxPol-1 \cite{hosseini2019encoding}} &0.4584 &0.5048 &0.5032 &0.6170 \\
		\textbf{HVS-MaxPol-2 \cite{hosseini2019encoding}} &0.5318 &0.4742 &0.2991 &0.5736 \\ \hline
		\textbf{Proposed ({\it SF} only)} &0.8273 &0.7964 &0.7766 &0.7867 \\
		\textbf{Proposed ({\it SN} only)} &0.6210 &0.5118 &0.4975 &0.6115 \\
		\textbf{Proposed SFSN} &\textbf{0.8979} &\textbf{0.8379} &\textbf{0.8004} &\textbf{0.8407} \\ \hline
	\end{tabular}
\label{table2}
\end{table}

Since different categories of SISR methods often generate drastically different appearance of the reconstructed images, it is intriguing to investigate how IQA methods perform for different SISR categories. The results on the QADS~\cite{zhou2019visual} database are reported in Table~\ref{table2}, where the proposed method delivers superior performance in each of the interpolation-based, dictionary-based, and DL-based SISR categories, as well as when all three categories are evaluated together. Ablation test has also been conducted to assess the performance when only the {\it SF} or {\it SN} measure is employed. The results are shown in Tables~\ref{table1} and \ref{table2}. It appears that both {\it SF} and {\it SN} measures make important contributions, but the best performance is achieved by the SFSN model that combines both of them.

\section{Conclusion}
In this work, we opt to a 2D approach to assess the quality of SISR generated images as a tradeoff between structural fidelity and statistical naturalness. This allows us to better understand the nature of quality degradations and better observe the varying behaviors of different SISR algorithms. We also show that a rather straightforward implementation of a local structural fidelity assessment, a global statistical naturalness measure, and a linear combination of the two, results in  an SFSN model that achieves surprisingly high correlations with MOS. In the future, better structural fidelity and statistical naturalness measures, and more sophisticated combination methods may be developed. The 2D assessment idea may also be integrated into novel SISR algorithms, aiming to achieve an optimal balance between the two goals.







\bibliographystyle{IEEEtran}
\bibliography{refs}
%
%
%

\end{document}